\documentclass{ws-ijmpa}
\usepackage{amssymb}
\usepackage{times}

\begin{document}

\markboth{E.~V.~Bugaev}
         {High Energy Neutrinos as a Probe for New Physics and Astrophysics}

\catchline{}{}{}{}{}

%\title{HIGH ENERGY NEUTRINOS: THEORY AND EXPERIMENTS}
% Changed according to referee's comment

\title{HIGH ENERGY NEUTRINOS AS A PROBE \\ FOR NEW PHYSICS AND ASTROPHYSICS}

\author{E.~V.~BUGAEV}
\address{Institute for Nuclear Research of Russian Academy of Sciences,
         Moscow 117312, Russia \\ bugaev@pcbai10.inr.ruhep.ru}

\maketitle

\pub{Received (Day Month Year)}{Revised (Day Month Year)}

\begin{abstract}
A review of the recent achievements in high energy neutrino physics and, 
partly, neutrino astrophysics is presented. It is argued that experiments with 
high energy neutrinos of natural origin can be used for a search of new physics 
effects beyond the electroweak scale.
\end{abstract}

\section{Introduction}

The typical theoretical prediction for an extragalactic neutrino flux at 
$E_{\nu}\sim10^{16}-10^{18}$~eV is the generic upper bound on the diffuse 
neutrino spectrum from AGN jets obtained by Mannheim {\it et~al.}\cite{1}.
In their model AGN source is not completely thin for a CR flux, and the 
normalization of the resulting CR spectrum is such that the CR intensity
does not exceed the proton spectrum obtained from observations.
At $E_{\nu}\sim10^{16}$~eV this bound is
\[
j(E)E^2<3\times10^{-7}~\mathrm{GeV}\cdot\mathrm{cm}^{-2}
\cdot\mathrm{sr}^{-1}\cdot\mathrm{s}^{-1}
\]
($j(E)$ is the differential diffuse neutrino spectrum). More conservative 
bound\cite{2} corresponds to the case when the sources are completely 
transparent for CRs: $j(E)E^2<2\times10^{-8}$ (independently on the neutrino 
energy if $j(E){\propto}E^{-2}$).

At higher energies, $E_{\nu}\sim10^{20}$~eV, extragalactic neutrino flux can 
have several components: the guaranteed component is the ``cosmogenic'' neutrino 
flux resulting from interactions of CRs with relic photons of CMB(cosmic 
microwave background). According to the most optimistic predictions\cite{3} the 
value $j(E)E^2$ for the cosmogenic flux may be as large as
\[
3\times10^{-7}~\mathrm{GeV}\cdot\mathrm{cm}^{-2}
\cdot\mathrm{sr}^{-1}\cdot\mathrm{s}^{-1}
\]
at $E_{\nu}\sim10^{20}$~eV. Other components of neutrino flux at ultrahigh energies 
are hypothetical: e.g., neutrino from decays of topological defects, neutrino 
from interactions of new (exotic) hadrons with mass $\sim(2-5)$~GeV in CR 
flux\cite{4} with relic photons etc.

Experimental bounds on the extragalactic diffuse neutrino flux are on the level 
\[
j(E)E^2\approx10^{-6}~\mathrm{GeV}\cdot\mathrm{cm}^{-2}
\cdot\mathrm{sr}^{-1}\cdot\mathrm{s}^{-1}
\] 
obtained, at $E\approx 10^{15}\mathrm{eV}$, in experiments on neutrino telescopes,
and, at higher energies,on large air shower arrays (see, e.g., review~\refcite{5}).

\section{Neutrino-nucleon interactions}
\label{sec:1}

\subsection{$\nu N$ DIS in Standard Model}

Generically, the differential cross section for the process $\nu_l N \to lX$ at 
large neutrino energies strongly depends on the behavior of parton distribution 
functions (pdfs) at small values of Bjorken variable $x$ and large values of 
$Q^2$. Besides, the gauge boson propagator effects are sufficient (leading to a 
nonlinear rise of $\sigma_{\nu N}(E_\nu)$) when $Q$ becomes comparable to the 
electroweak scale, $Q^2\gg M^2_W$. In the energy region which we are interested 
in, $E_{\nu}>10^7$~GeV, one has
\[
Q^2_{\mathrm{char}} \approx M^2_W, \quad x_{\mathrm{char}} < 10^{-4}-10^{-5}.
\]

In QCD-improved parton model perturbative QCD corrections to the nonperturbative 
pdfs (taken from experiment) are calculated using two approaches: DGLAP scheme 
(resummations of the perturbative expansion retaining leading terms in 
$\ln(\frac{Q^2}{\Lambda^2_{\mathrm{QCD}}})$) and BFKL scheme (resummations 
retaining leading terms in $\ln(\frac{1}{x})$). Both these logarithms become 
large in the relevant kinematic region. Since, very roughly, $x$-dependence of 
pdfs of sea quarks has a power-law behavior, $\propto x^{-0.3}$, for $x\ll1$,
it follows that
\[
\sigma_{\nu N}(E_\nu) \propto E_\nu^{0.3}.
\]
The power-law rise of the cross section implies a violation of unitarity at very
high energies.

It is rather evident that there must be non-linear (``higher twist'') QCD 
corrections to the standard calculation, those which are beyond the linear 
evolution DGLAP and BFKL schemes. Physically, these non-linear corrections arise 
due to the growth of parton (gluon) density in the nucleon target at small $x$. 
Taking into account the recombination gluon effects leads to a taming of the 
fast rise of the cross section, in agreement with unitarity requirements.

There are different approaches accounting these gluon screening effects. To feel 
the order of magnitude of corresponding corrections to $\sigma_{\nu N}(E_{\nu})$ 
it is enough to compare the results of two models: unified BFKL/DGLAP + 
screening model\cite{6} and phenomenological colour-dipole model.\cite{7}
In the latter model, the DIS is considered in the laboratory system: virtual
weak gauge boson at some distance from the target fluctuates into $qq$-pair
interacting with the target via multiple gluon exchanges. This approach,
in its original form, does not take into account QCD evolution at all.
It appears that at extremely high neutrino energies,
$E_\nu\approx10^{12}~\mathrm{GeV}$, the  difference in the predictions
of these models for the total cross section does 
not exceed factor $2-3$ (see Ref.~\refcite{6}).

The $\sigma_{\nu N}(E_\nu)$ scales, approximately, as $E_{\nu}^{0.363}$ for 
$E_{\nu}>10^{16}$~eV. Using this one can easily obtain the estimate for the 
rate of contained events in neutrino telescopes.\cite{8} According to this 
estimate, the value $j(E)E^2$ should be larger than
\[
10^{-7}$~$\mathrm{GeV}\cdot\mathrm{cm}^{-2}\cdot\mathrm{sr}^{-1}\cdot\mathrm{s}^{-1}
\] 
to be measurable at $E_{\nu}>10^{19}$~eV at $1\;\mathrm{km}^3$ detector, if the
SM prediction for the total cross section is correct.

\subsection{TeV scale gravity models for $\sigma_{\nu N}$}

It is easy to show that the gravitational scattering of any two particles
(via graviton exchange) becomes strong if
$\sqrt{s}\gg M_{\mathrm{Pl}}\approx10^{19}$~GeV, i.e., at inaccessibly high 
energies of colliding particles.

In TeV-scale quantum gravity models proposed by Arkani-Hamed
{\it et~al.}\cite{9} there are two most important independent parameters: $M_D$ 
(fundamental gravitational, or Planck, scale) and $n$ -- the number of extra 
(compact) dimensions. The value $M_D$ is connected with the 4-dimensional Newton 
constant through the compactification radius $R$. If the compactification 
volume, $V=2\pi R$, is large, $M_D$, which is equal to 
$\left(\frac{M_{\mathrm{Pl}}^2}{V^n}\right)^{\frac{1}{n+2}}$, is much smaller 
than $M_{\mathrm{Pl}}$. In TeV-scale gravity models one has $M_D\approx1$~TeV.

In these models, because of the compactification, the extra $n$ components of 
the graviton momentum are quantized. To an observer in the usual 4-dimensional 
space-time, the graviton would appear to be a massive Kaluza-Klein (KK) state. 
The result of the summation over KK states in the expression for the amplitude 
of gravitational scattering is
\[
A\approx\frac{s^2}{M_D^4}\left(\frac{\Lambda}{M_D}\right)^{n-2},
\]
where $\Lambda(\approx M_D)$ is the cut-off constant (the Born amplitude is 
divergent but after eikonalization the amplitude becomes finite). We see from 
this expression that the amplitude is large if
$\sqrt{s}\ge M_D\approx1\mathrm{TeV}$, rather than $\sqrt{s}>M_{\mathrm{Pl}}$, 
as in the 4-dimensional theory. Due to $s^2$-dependence, the gravitational 
scattering amplitude becomes, at some energy, larger than any other amplitude.

The gravitational amplitude and cross section can be calculated reliably if the 
process is semiclassical, i.e., when the Schwarzschild radius ($R_S$) of the 
colliding system (which is the classic value) is much larger than quantum 
lengths, $\lambda_p$ (it is higher dimensional Planck length) and $\lambda_B$ 
(de Broglie wave length of the colliding particles). It is easy to see, that for 
the semiclassical picture of the gravitational scattering the inequality 
$\sqrt{s}\gg M_D$ must take place (the ``trans-Planckian energy regime''). In 
this regime gravitational interaction dominates over other gauge interactions 
and is a semiclassical process.

The important characteristic of the gravitational scattering is the impact 
parameter $b$. If $b\gg R_S$ one has $-\frac{t}{s}\ll 1$ ($t$ is the square of 
the momentum transfer) and it corresponds to a small scattering angle limit
(and elastic scattering). If, in opposite case, $b<R_S$, the collision leads
to a production of a black hole with Schwarzschild radius $R_S$ and mass 
$M_{\mathrm{BH}}$ equal, roughly, to cms energy $\sqrt{s}$.

The condition $\sqrt{s}\gg M_D$ corresponds to the existence of the minimum 
value of BH mass which can be reliably produced in the collision:
\[
M_{\mathrm{BH}}^{\min}=\sqrt{s_{\min}}=\alpha M_D,\quad \alpha\gg1,
\]
$\alpha$ is the model parameter. The cross section of the BH production in
${\nu}N$ collision is equal to
\[
\sigma_{\nu N}^{\mathrm{BH}}=\sum_{i}\int\limits_0^1 dx f_i (x,\mu)
\sigma_{\nu i}^{\mathrm{BH}}(xs),
\]
where $f_i(x,\mu)$ is pdf for the parton of type $i$, $\mu\approx R_S^{-1}$,
and $\sigma_{\nu i}^{BH}$ is the cross section of BH production in $\nu$-parton 
collision at cms energy $\sqrt{xs}$,
\[
\sigma_{\nu i}^{\mathrm{BH}}(xs)\approx \pi R_{S}^2 (M_{\mathrm{BH}}=\sqrt{xs})
\Theta (\sqrt{xs}-\alpha M_D).
\] 

The differential elastic gravitational scattering of neutrino on nucleon, 
$\frac{d\sigma}{dy}$, is calculated (using multiple graviton exchange\cite{10}) 
as a function of $y=\frac{E_{\nu}-{E_{\nu}}'}{E_{\nu}}$. This cross section 
grows as $y$ decreases. The small $y$ region corresponds to long distance 
processes where neutrino interacts with a parton and transfers only a small 
portion of its energy, surviving after interaction.

\subsection{Neutrino interactions in string theory}

If cms energies of collisions are close to the fundamental scale of gravity, 
i.e., if $\sqrt{s}\approx M_D$ (``Planckian region''), the classical description 
cannot be trusted. String theory provides the best hope for understanding the 
regime of strong quantum gravity, and for computing amplitudes at energies close 
to $M_D$. So, the models with extra dimensions must be embedded in realistic 
string models, in which the unification of gravity with the SM takes place.

In string theory the massless graviton is the zero mode of a closed string, 
whereas the gauge bosons of SM are the lightest modes of an open string. The 
scattering amplitudes in string theory are amplitudes of string exchanges and 
are described by the formulas of Veneziano's (exchange of an open string) or 
Virasoro's (exchange of a closed string) type. Correspondingly, one expects the 
presence of string Regge (SR) excitations of the graviton and gauge bosons 
(analogously to resonances in Veneziano model). By duality arguments, if there 
are amplitudes with SR excitations exchanges in the $t$-channel, having 
corresponding $t$-channel poles, these amplitudes will also have $s$-channel 
poles.

In lepton-quark scattering the $s$-channel resonances are leptoquarks.
From the resonance amplitude 
\[
\nu + q \to X_n^J \to \nu + q
\]
($J$ is a spin of the resonance) one can obtain the cross section,\cite{ins1}
$\sigma_n^J (\nu q)\equiv\sigma\left(\nu q \to X_n^J\right)$, and in the narrow
width approximation one has
\[
\sigma_n^J (\nu q)\sim g^2 \delta\left(s-nM_S^2\right),
\]
where $g$ is the gauge coupling constant and $M_S$ is the string scale. 
One can assume that $M_S\sim M_D\sim1$~TeV, and, in this case, the SR resonances 
are ``TeV strings''. So, we have, in fact, the GUT unification at TeV scale.

\subsection{Search for a new physics}

Neutrino experiments with neutrino of natural origin can be effectively used
for a search of a physics beyond the standard model. There are different 
possibilities.
\begin{enumerate}
\item[1.] {\it Study of contained events in large neutrino telescopes.}
          At large enough energies of neutrino, say,
          ${E_{\nu}>500-1000~\mathrm{TeV}}$, most part of such events may be
          due to gravitational elastic scattering of neutrino on the nucleon
          target or BH production by neutrinos. One can study the angular
          dependence, up-down ratio as well as the energetic distribution.\cite{11}
\item[2.] {\it Study of through-going muons with ${E>E_0}$, produced in BH 
               production processes in large neutrino telescopes.\cite{12}}
\item[3.] {\it Study of quasi-horizontal air showers which occur at rates 
               exceeding the predictions of SM and have distinct characteristics} 
               (see, e.g., Ref.~\refcite{13}).
          Although greatly reduced by hypothetical BH production process (the 
          cross section of which can be, in TeV gravity models, about $10^4$~nb 
          at $10^{20}~\mathrm{eV}$), neutrino interaction lengths, 
          $L=1.7\times10^7$~km~w.e. ($\frac{\mathrm{pb}}{\sigma})$, are still
          far larger than the Earth's atmospheric depth, which is $L=0.36$~km~w.e.,
          when traversed horizontally. Neutrinos therefore produce black holes 
          uniformly at all atmospheric depths. As a result, the most promising 
          signal of BH production by CRs is quasi-horizontal showers initiated 
          by neutrinos deep in atmosphere. Distinct characteristics of the 
          showers are following: anomalous electromagnetic component;\cite{14} 
          large multiplicity; large $\frac{\mu}{e}$ ratio; they look
          nucleus-like (for not a small $X_{\max}$); they have a curved front,
          with particles well spread in time.
\item[4.] {\it Earth-skimming idea}.
          The large detectors of air showers can register the fluxes of UHE 
          neutrinos and, simultaneously, will be able to measure 
          ${\sigma_{{\nu}N}}$ at energies as high as\cite{15} 
          $10^{20}~\mathrm{eV}$. Several proposed experiments plan to detect UHE 
          neutrinos by observation of nearly horizontal air showers (HAS) 
          resulting in atmosphere from $\nu$-air interactions.
          At $E\approx10^{20}~\mathrm{eV}$ the ${\sigma_{\nu N}}$ in SM is about 
          $10^{-31}~\mathrm{cm}^2$. The air shower probability of HAS production 
          is proportional to ${\sigma_{\nu N}}$. In addition to HAS, the 
          experiments can also observe up-going air showers (UAS) initiated by 
          muon and tau leptons produced by neutrinos interacting just below the 
          surface of the Earth. The expected rate of UAS depend non-linearly on 
          the cross section (if $\sigma_{\nu N}$ larger than $10^{-32}$~cm$^2$, 
          UAS probability is inversely proportional to $\sigma_{\nu N}$). Thus, 
          by comparing the HAS and UAS rates, the cross section can be 
          determined. Taken together, these rates ensure a total event rate, 
          weakly depending of the value of $\sigma_{\nu N}$. If there is a new 
          physics (e.g., BH production), this idea is also useful. Really, a 
          large rate of quasi-horizontal showers may be attributed to either an 
          enhanced neutrino flux or an enhanced BH cross section. While an 
          enhanced flux increases these rates, a large BH cross section will 
          suppress them, since the hadronic decay products of BH evaporation 
          will not escape the Earth's crust.\cite{13}
\end{enumerate}

\section{Neutrinos and UHECR puzzle}
\label{sec:3}

\subsection{$Z$-burst model}

It had been proposed\cite{16} that the primaries which propagate across 
distances above the GZK zone ($\approx50$~Mpc) are neutrinos, which then 
annihilate with relic neutrinos within this zone to create a flux of nucleons 
and photons with energies above $E_{\mathrm{GZK}}$. The annihilation cross 
section is relatively large ($\approx10^{-5}$~mb) near $Z$-resonance and the 
resonance neutrino energy depends on the neutrino mass.

There are three main difficulties of the $Z$-burst model.
\begin{enumerate}
\item[1.] Primary protons have to be accelerated to extremely high energies, 
          $E>10^{23}$~eV, in order to produce in astrophysical sources (via $pp$ 
          and $p\gamma$ reactions) UHE neutrinos. The photons produced in the 
          same reactions have to be absorbed inside the source otherwise the 
          diffuse background of MeV--GeV photons will be too large. Hidden 
          sources are unable to produce neutrinos of such high energies because 
          the decay length of pions becomes equal to its scattering length. The 
          luminosities of sources required in this model are too high,\cite{17} 
          $\approx(10^{45}-10^{47})$~erg/s.
\item[2.] Even if one assumes that neutrinos originate from decays of superheavy 
          dark matter particles through the only channel, $X-\nu\tilde{\nu}$, 
          the problems with diffuse photon background are not avoided because 
          higher order corrections to this process give rise to electroweak 
          cascades transferring large fraction of energy to photons and 
          electrons.\cite{17}
\item[3.] The required UHE neutrino fluxes in the $Z$-burst model are very 
          large:\cite{3} they are almost excluded by the new experimental limits 
          from FORTE and GLUE experiments (as well as by the new limit from 
          EGRET\cite{18}).
\end{enumerate}

It was shown in the work of Gelmini {\it et~al.}\cite{19}, using
the AGASA data, that the nonobservation of CR events at 
$E^{\mathrm{CR}}>2\times10^{20}~\mathrm{eV}$ implies a lower bound 
$\approx0.3~\mathrm{eV}$ on the neutrino mass. Since this value exceeds 
$\sqrt{\Delta m^2_{ij}}$ from neutrino oscillation experiments, the bound 
applies to all three neutrino masses. If there is neutrino mass hierarchy then 
it follows from SK data that the mass of the heavier neutrino is 
$\approx0.04~\mathrm{eV}$. It is argued by Gelmini {\it et~al.}, that AGASA data 
are incompatible with such a low value because in this case the resonance energy 
is too high and this predicts too many CRs beyond the AGASA end point, 
$2\times10^{20}~\mathrm{eV}$. So, the bound leaves only a small interval for 
neutrino mass around $\approx0.3~\mathrm{eV}$ (if $Z$-burst model is correct).

The flux requirement obtained in the work of Gelmini {\it et~al.}, is about 
$7\times10^{-36}~(\mathrm{eV}\cdot\mathrm{m}^2\cdot\mathrm{sr}\cdot\mathrm{s})^{-1}$,
and such a high value is marginally excluded by GLUE and FORTE experiments.
The relic neutrino density could be enhanced by either gravitational clustering
or by a large lepton asymmetry. Calculations argue against of such a clustering
of relic neutrinos if neutrino mass is so small ($m_\nu\approx0.3~\mathrm{eV}$):
the overdensity $\delta$ of neutrinos in our Local Group of galaxies is $<10$,
on a length scales $\ge1\mathrm{Mpc}$ (see Ref.~\refcite{20}). Also, BBN physics 
gives the bound on a lepton asymmetry, $\left|\xi\nu_i\right|<0.1$, from which
it follows that the extra contribution to $\rho_\nu$ from degeneracy is very 
small.\cite{21}

\subsection{Neutrinos as CR primaries}

It was noted above that within the SM the neutrino-nucleon interaction cross 
section at $10^{20}~\mathrm{eV}$ is about $10^{-4}~\mathrm{mb}$, i.e., on five 
orders of magnitude smaller than necessary to produce air showers starting high 
in the atmosphere. It appears also that models with extra dimensions predict 
cross sections which on a factor $\approx10^2$ larger than that in SM. It is 
clearly not enough for explanation of the GZK puzzle.

It was argued recently\cite{22} that the cross section at $E>10^{18}$~eV can be 
greatly (on a factor $10^5-10^6$) enhanced by nonperturbative electroweak 
instanton contributions. The processes induced by electroweak instantons
(which represent tunneling transitions between topologically inequivalent vacua) 
violate baryon + lepton (B+L) number and are characterized by the large 
multiplicity of final state particles (quarks, gauge and Higgs bosons).The 
transition rate of the tunneling process is exponentially suppressed at low 
energies when
\[
E_{\mathrm{CMS}}\ll4E_{\mathrm{sph}}\approx4\pi\frac{M_W}{\alpha_W}\approx30~\mathrm{TeV}.
\]
Here, $E_{\mathrm{sph}}\approx8~\mathrm{TeV}$ is the sphaleron energy. At high 
temperatures and high energies the transition rate can be unsuppressed, leading 
to observable effects in CR experiments. According to Ref.~\refcite{22}, at 
$E_{\mathrm{lab}}\approx10^{20}~\mathrm{eV}$ the cross section is about $3$~mb.

According to the calculation of Bezrukov {\it et~al.}\cite{23} based on a 
generalized semi-classical approach, a severe exponential suppression of the 
cross section is extended up to energy 
$30E_{\mathrm{sph}}\approx250~\mathrm{TeV}$. Bezrukov {\it et~al.} did not 
calculate the cross section, but estimated only the {\em upper limit} of the 
exponential function. Therefore one cannot obtain the reliable cross section 
from their paper although it is tempting to do this.\cite{24}

\section{Neutrino from dark matter}
\label{sec:4}

The existence of dark matter (DM) is the experimental evidence for new physics 
beyond the SM. The most recent WMAP data\cite{ins2} give the amount of cold dark 
matter (CDM) as
\[
0.095<\Omega_{\mathrm{CDM}}h^2<0.129, \quad (2\sigma~\mathrm{C.L.})
\]
and the total matter density
\[
0.126<\Omega_m h^2<0.143.
\]

There are many particle dark matter candidates for a stable WIMPs (weakly 
interacting massive particles) which could have the relevant relic density.

Indirect DM searches have been proposed\cite{25} to observe the products of DM 
annihilation including neutrinos. WIMPs, which scatter elastically in the Sun or 
Earth, may become gravitationally bound and, over the age of the solar system, 
they may accumulate in these objects, greatly enhancing their annihilation rate. 
Neutrinos can escape the Sun or Earth and can be registered in a detector.

\subsection{SUSY DM}

All superpartners are charged under a discrete symmetry called $R$-parity
($+1$ for SM particles and $-1$ for their superpartners). $R$-parity conservation
guarantees that the lightest superpartner, being odd under $R$-parity, is
absolutely stable and becomes a DM candidate.

The desired WIMP is the lightest neutralino $\tilde\chi_1^0$ which is a mixture 
of the superpartners of the hypercharge gauge boson ($\tilde{b}^0$), the neutral 
$SU(2)_W$ gauge boson ($\tilde{w}^0$) and the two neutral Higgs bosons,
\[
\tilde\chi_1^0 = a_1\tilde{b}^0   + a_2\tilde{w}^0
                +a_3\tilde{h}_u^0 + a_4\tilde{h}_d^0.
\]
The recent progress in particle physics and cosmology favors the region of the 
SUSY theory parameters where the neutralino is a Bino-Higgsino mixture rather 
than Bino-like.

The rate of neutrino production in WIMP annihilations is highly model dependent. 
Neutralinos are Majorana fermions, and the Pauli exclusion principle suppresses 
the direct production in the annihilations of pairs of light fermions, so only 
indirect channels of neutrino production are available, through the decay of 
particles produced in the processes
\[
\chi\chi\to\tau^+\tau^-, b\tilde{b}, W^+W^-, ZZ, HH, t\tilde{t}.
\]
If neutralino is Bino-like, the production of $SU(2)_W$ gauge bosons is 
suppressed. 

If neutralino exists and is heavy enough and if it is really a mixture
Bino-Higgsino, the neutrino signal from the Sun direction can be detected by
large neutrino telescopes (see, e.g., Ref.~\cite{26}). The limits on neutralino
mass from experiments and from calculations of relic number and mass density
are very weak:\cite{27}
\[
20~\mathrm{GeV}\lesssim{m}_{\chi}\le600~\mathrm{GeV}.
\]

\subsection{Kaluza-Klein DM}

There is one class of models with extra dimensions, in which all of the fields 
of the SM propagate in these dimensions, not only gravitons. Each SM field has 
an infinite tower of KK partners with identical spins and couplings and masses 
of order n/R. Naturally, in such models the conservation of the momentum along 
the extra dimensions and, as a consequence, the conservation of KK number, takes 
place. Radiative corrections break KK number down to a discrete conserved 
quantity, KK-parity: $(-1)^n$, where $n$ is the number of the KK level.
KK-parity ensures that the lightest KK-partner (LKP) at level one, being
odd under KK-parity, is stable, and can be a DM candidate.

In the concrete model\cite{28} the LKP is a neutral WIMP which is a linear 
combination of the first KK mode $B_1$ of the hypercharge gauge boson and the 
first KK mode $W^0_1$ of the neutral $SU(2)_W$ gauge boson.

The calculation of the relic density of KK DM based on WMAP's results for 
$\Omega_{\mathrm{CDM}}$ predicts the value of LKP mass:\cite{29} 
\[
m_{\mathrm{LKP}}\approx(600-1200)~\mathrm{GeV}.
\]

In contrast with LSP, LKP is a vector particle, and the helicity suppression is 
absent, so, LKP can directly annihilate into neutrinos. Besides, the 
annihilation into pairs of top-quarks is sufficient due to a large value of LKP 
mass. Top-quarks almost always decay in the channel $t\to\mathrm{Wb}$. The $W$,
in turn, decays with equal branching ratios into $\nu_ee$, $\nu_{\mu}\mu$, 
$\nu_{\tau}\tau$. It is important that both decays, $t$'s and $W$'s (and also 
$\tau$'s) take place without any energy loss. So,in this model, the Sun is 
effective source of high energy ($E_{\nu}\approx100$~GeV--$1$~TeV) tau-neutrinos 
and (due to oscillations) muon neutrinos.

\subsection{Superheavy DM}

It is well known that the assumption that DM is a thermal relic is too 
restrictive. Unitarity bound on the annihilation cross section and the 
assumption of thermal equilibrium in the early universe lead to the bound\cite{30}
\[
\Omega h^2 \ge 0.1\left(M/10^5~\mathrm{GeV}\right)^2,
\]
where $\Omega$ is the ratio of the DM mass density to $\rho_c$ today. However,
if WIMPs are not thermal relics, their masses are not thermodynamically 
constrained, i.e., they may be even superheavy (``wimpzillas''). Moreover,
their present day abundance does not depend on whether they have strong 
(``simpzillas''), weak, electromagnetic, or only gravitational interactions. 

Such particles can be, e.g., gravitationally produced near the end of inflation, 
due to non-adiabatic change of the scale factor during the transition from the 
de-Sitter to the radiation dominated phase.\cite{31}

High energy neutrinos are produced by the simpzilla annihilations, which produce 
a quark or gluon pair which then fragment into hadronic jets. Neutrino fluxes 
from the Sun were calculated by Albuquerque {\it et~al.}\cite{32}. It was shown 
that $\nu_\tau$ and $\nu_{\mu}$ fluxes can be rather large and measurable by 
neutrino telescopes, with $E_{\nu}$ in the interval ($50$~GeV -- $1$~TeV),
the spectrum is very similar with the case of KK DM (the channel $t\to{Wb}$ 
also works). One should note that the typical $E_{\nu}$ value in the spectrum 
is much lower than the simpzilla mass ($M\approx10^8-10^{16}$~GeV), due to the 
large numbers of neutrinos produced per annihilation.

Unfortunately,direct dark matter search experiments exclude\cite{33} the most 
natural values of simpzilla mass (those which are comparable with the inflaton 
mass in chaotic inflation models ($\approx10^{12}$~GeV)).

\end{document}